\documentclass{JINST}

\usepackage{graphicx}
\usepackage{xspace}
\usepackage{amsmath,amssymb,epsfig}

\newcommand{\GeVc}{\ensuremath{\mbox{GeV}/c}\xspace}
\newcommand{\MeVc}{\ensuremath{\mbox{MeV}/c}\xspace}

\newcommand{\mm}{\ensuremath{\mbox{mm}}\xspace}

\newcommand{\ps}{\ensuremath{\mbox{ps}}\xspace}

\title{The Time Response of Glass Resistive Plate
  Chambers to Heavily Ionizing Particles}

\author{%
A.~Artamonov$^a$, 
A.~Blondel$^b$, 
M.~Bogomilov$^c$,
C.~Booth$^d$,
S.~Borghi$^{b,1}$, 
M.~G.~Catanesi$^e$,
A.~Cervera--Villanueva$^f$,
P.~Chimenti$^g$,
U.~Gastaldi$^h$, 
S.~Giani$^i$,
J.J.~G\'{o}mez--Cadenas$^f$,
J.S.~Graulich$^{j,2}$,
G.~Gr\'{e}goire$^j$,
A.~Grossheim$^{h,3}$,
A.~Guglielmi$^k$,
V.~Ivanchenko$^{h,4}$, 
D.~Kolev$^c$,
C.~Meurer$^l$, 
M.~Mezzetto$^k$, 
J.~Panman$^h$, 
B.~Popov$^m$, 
E.~Radicioni$^e$,
R.~Schroeter$^b$,
P.~Temnikov$^n$, 
E.~Tcherniaev$^h$,
R.~Tsenov$^c$\thanks{Corresponding author, e-mail: Roumen.Tsenov@cern.ch.}~, 
I.~Tsukerman$^a$
and
C.~Wiebusch$^o$\\
\llap{$^a$}%
{ITEP, Moscow, Russian Federation}\\
\llap{$^b$}%
{Section de Physique, Universit\'{e} de Gen\`{e}ve, Switzerland}\\
\llap{$^c$}%
{Faculty of Physics, St. Kliment Ohridski University of Sofia, Bulgaria}\\
\llap{$^d$}%
{Dept. of Physics, University of Sheffield, UK}\\
\llap{$^e$}%
{Universit\`{a} degli Studi e Sezione INFN, Bari, Italy}\\
\llap{$^f$}%
{Instituto de F\'{i}sica Corpuscular, IFIC, CSIC and Universidad de Valencia,
Spain}%
\llap{$^g$}%
{Universit\`{a} degli Studi e Sezione INFN, Trieste, Italy}\\
\llap{$^h$}%
{Laboratori Nazionali di Legnaro dell' INFN, Legnaro, Italy}\\
\llap{$^i$}%
{CERN, Geneva, Switzerland}\\ 
\llap{$^j$}%
{Institut de Physique Nucl\'{e}aire, UCL, Louvain-la-Neuve,
  Belgium}\\
\llap{$^k$}%
{Universit\`{a} degli Studi e Sezione INFN, Padova, Italy}\\
\llap{$^l$}%
{Institut f\"{u}r Physik, Forschungszentrum Karlsruhe, Germany}\\
\llap{$^m$}%
{Joint Institute for Nuclear Research, JINR, Dubna, Russia}\\
\llap{$^n$}%
{Institute for Nuclear Research and Nuclear Energy, 
Academy of Sciences, Sofia, Bulgaria}\\
\llap{$^o$}%
{III Phys. Inst. B, RWTH Aachen, Aachen, Germany}\\
\llap{$^1$}%
{Now at CERN, Geneva, Switzerland}\\
\llap{$^2$}%
{Now at Section de Physique, Universit\'{e} de Gen\`{e}ve, Switzerland}\\
\llap{$^3$}%
{Now at TRIUMF, Vancouver, Canada}\\
\llap{$^4$}%
{On leave of absence from Ecoanalitica, Moscow State University,
  Moscow, Russia}\\ 
}
 
\abstract{
The HARP system of resistive plate chambers (RPCs) was designed to
perform particle identification by 
the measurement of the difference in the time-of-flight of different
particles. 
In previous papers an apparent discrepancy was shown between the
response of the RPCs to minimum ionizing pions and heavily ionizing
protons. 
Using the kinematics of elastic scattering off a hydrogen target a
controlled beam of low momentum recoil protons was directed onto the RPC
chambers. 
With this method the trajectory and momentum, and hence the
time-of-flight of the protons can be precisely predicted without need
for a measurement of momentum of the protons.
It is demonstrated that the measurement of the time-of-arrival of
particles by the thin gas-gap glass RPC system of the HARP experiment
depends on the primary ionization deposited by the particle in the
detector. 
}


\begin{document}

\section{Introduction}
\label{sec:intro}

The HARP experiment~\cite{harpprop}, \cite{harpnim} has been designed to measure
 hadron production cross-sections on nuclear targets with a precision of
 a few percent over almost the full solid angle. 
A set of solid and liquid targets spanning a large range in
 atomic number was exposed to beams of protons and pions with momenta
 between 1.5~\GeVc and 15~\GeVc. 
The elements used ranged from hydrogen to lead.
HARP took 450 million physics triggers, collected data 
for about 300 different settings and recorded 30 TB of information 
from August 2001 to October 2002.

\begin{figure}
\centering
\includegraphics[width=4.in]{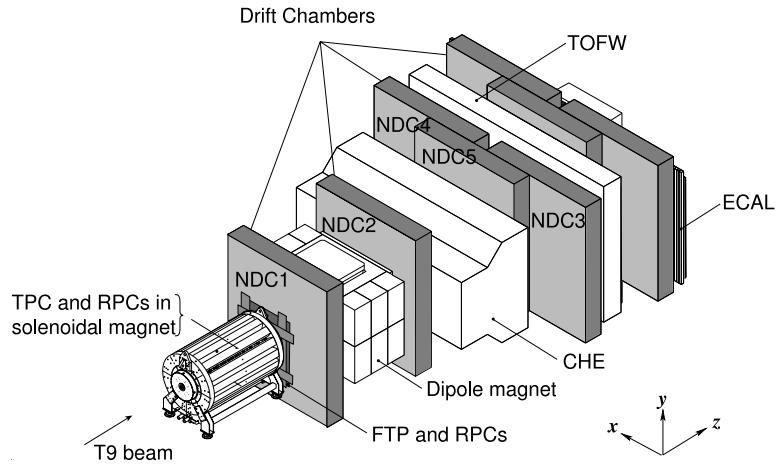}
\caption[] {\small {Schematic layout of the HARP detector. The detector
covers a total length of 13.5~m along the 
beam direction and has a maximum width of
6.5~m perpendicular to the beam.}}
\label{detector}
\end{figure}

The setup of the HARP experiment is shown in figure~\ref{detector}. 
A detailed 
description of the detector and its performance is given in
Ref.~\cite{harpprop}, \cite{harpnim}. 
The spectrometer can be subdivided into three main systems.  
{\textit{Beam and trigger detectors}} provide tracking and
  identification of beam particles, and trigger decisions.
{\textit{Forward detectors}} provide tracking, momentum
  measurement and identification of secondary
particles at angles less than $17^\circ$ with respect to the beam axis.
{\textit{Large angle detectors}} deal with tracking, momentum
  measurement and particle identification at large production angles.

The large angle detector system consists of 
\textit{Time Projection Chamber (TPC)},
\textit{Resistive Plate Chambers (RPC)} and  
 an \textit{Inner Trigger Cylinder (ITC)}.

\begin{figure}[tbp]
\vspace{9pt}
\begin{center}
  \epsfig{figure=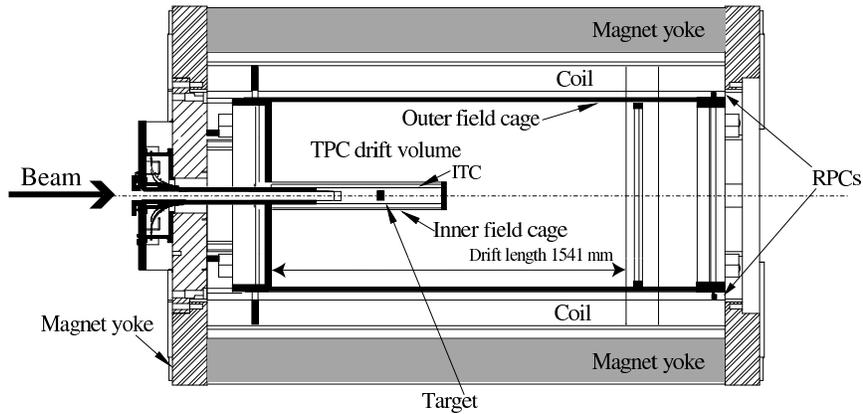,width=0.75\textwidth,angle=0}
\end{center}
\caption[] {\small {Schematic layout of the TPC. The
beam enters from the left. Starting from the
outside, first the return yoke of the magnet
is seen, closed with an end-cap at the upstream
end, and open at the downstream end.
The cylindrical coils are drawn inside the yoke.
The field cage is positioned inside this magnetic
volume. The inner field cage is visible
as a short cylinder entering from the left. The
ITC trigger counter and the target holder are
inserted in the inner field cage. The RPCs (not drawn)
  are positioned between 
the outer field cage and the coil, as indicated.}}
\label{tpc}
\end{figure}

The TPC has a cylindrical
form with a length 
of $\sim$~200~cm and a diameter of  $\sim$~83~cm. It is located inside
a solenoid magnet 
producing a 0.7~T field. It  measures momentum, trajectory 
and ionization energy losses of particles emitted from the target at 
large angles ($20^\circ < \theta < 160^\circ$) with respect to the
incoming beam. 
The RPCs are arranged in the shape of a barrel around the outer field cage
of the TPC. 
A target station and the ITC are situated
inside TPC volume, in a truncated inner field cage (see figure~\ref{tpc}). 

This paper concentrates on a specific effect found when calibrating
the time response of the HARP RPCs.
Besides the well-known {\it time-walk} correction, their time
response was found  {\em to depend
on the primary ionization deposited in the gas gap}. Brief information
needed to understand the RPC hardware and calibration procedures is given in
the next section. The evidence for the effect itself is given in the
Section~\ref{sec:effect}, 
followed by some discussion (Section~\ref{sec:discus}) and conclusions
(Section~\ref{sec:concl}).

\section{HARP RPC system}
\label{sec:hardware}
Particle identification  (PID) at large angles is mainly performed by 
mean energy loss ($dE/dx$) measurements in the TPC.
A complementary system for particle identification
to 
distinguish between electrons and pions in the momentum range
125~\MeVc--250~\MeVc is needed. For these relatively low momenta PID
can be 
performed by time-of-flight (ToF) measurements. 
A time resolution of $\sim$~200~\ps is
required when particles traverse the TPC over
the shortest distance ($\sim$~400~mm). 
ToF measurements with such a resolution also help in pion--proton
separation up to $\sim$~1~\GeVc.
A system of resistive plate chambers has been designed and
used as additional PID
device for particles emerging with large production angles. 

The design of the HARP RPCs is based on a  prototype developed 
for the ALICE experiment~\cite{alicerpc}, \cite{protvino_rpc}. 

\begin{figure}
\centering
\includegraphics[width=5.in]{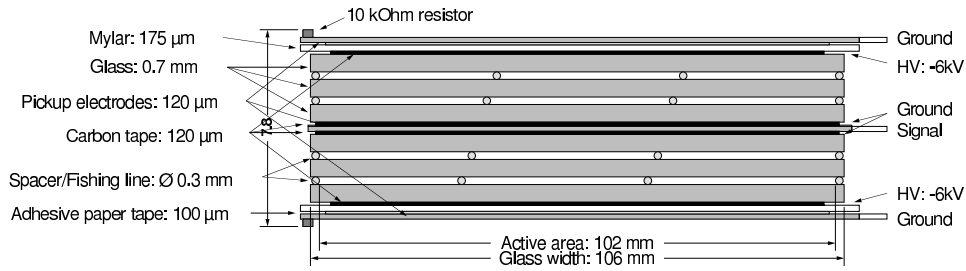}
\caption[] {\small{Cross-section of RPC glass stack. Six glass plates
form four 0.3~mm gaps using fishing lines as spacer.
 The only signal electrode is situated in the middle of the stack.}}
\label{rpc_stack}
\end{figure}

The RPCs are constructed as a four-gap
stack of glass plates. The gap size is precisely set 
to 0.3~mm by interposing a fishing line (nylon mono-filament) 
with suitable diameter between the plates.
The stack consists of two identical structures of three glass
plates each, arranged symmetrically on both sides of  a central
readout electrode. The glass plates are 0.7~mm thick and made of
standard float glass with a specific resistivity of $\sim 10^{13}~\Omega$cm. 
A view of the cross-section through the short side of the glass stack
is shown  
in figure~\ref{rpc_stack}. 
The negative high voltage is applied to the glass plates by means of a
coated graphite layer with resistivity of 200 $k\Omega/\Box$
on the two outer glass plates of both sets. A single readout
electrode, located in the center of the glass stack, 
collects signals from all four gaps. It is segmented into 
64 rectangular strips. Eight strips 
are connected to one pre-amplifier forming a readout channel. 
Thus, each chamber has eight readout channels, in the following
referred to as pads, and numbered from 1 to 8 starting from the most
upstream one. Pads with one and the same number define a pad-ring.
A chamber is 1920~mm long, 104~mm wide and 
7.8~mm thick. Each chamber is housed in an aluminum box with 
dimensions 2~m $\times$ 150~mm $\times$ 10~mm.

Thirty such chambers are 
arranged around the TPC in two staggered layers forming a barrel. They
cover polar  
angles from $17^\circ$ to $142^\circ$ with respect to the beam axis
and $2\pi$ in azimuthal angle with a small 13~mm overlap between the layers.
The set of two layers fits into a $\sim$ 25~mm radial space 
between the TPC and the coils of the solenoid magnet. 
The readout electrodes of the inner and outer 
layer are located at radial distances of 421~mm and 
434.5~mm, respectively. 
The front-end electronics of each channel consists of 
a pre-amplifier board mounted on the chamber and a combined
splitter/discriminator module. 

Pre-amplifiers are based on  the AD8009 chip operating with an
amplification  
factor of $\sim$ 30. The amplified signals are transmitted 
through mini coaxial cables over a distance of 0.8~m -- 2.5~m (depending
on the channel in question) to a 
passive 
patch panel
and from there over a distance of 5~m 
through Lemo 50~$\Omega$ cables to a custom-made 
 splitter and leading edge discriminator module. 
Each signal is amplified once more there
(with a factor $\sim$~3 using the same AD8009 chip) and split into
two separate signals.  One of them is discriminated (the discriminator
threshold is put at 5~mV) and is  
sent via a 80~m twisted pair cable to a Time--to--Digital Converter 
(TDC), model CAEN V775 with a channel nominal width of 35~\ps. 
The other signal is sent through another 80~m
twisted pair cable to a Charge--To--Digital Converter 
(QDC), model CAEN V792.

The RPCs were operated in avalanche mode at a voltage of $-6$~kV between
outer and central electrodes and with a
gas mixture of 90\% $\mbox{C}_2 \mbox{F}_4 \mbox{H}_2$, 5\%
$\mbox{SF}_6$, and 5\% $\mbox{C}_4 \mbox{H}_{10}$.  

Typical random noise rates were in the range of 200~Hz--300~Hz per
chamber, i.e. $\leq$ 1 kHz/m$^2$, which is an acceptable level compared 
to the typical particle rate 10 kHz/m$^2$ over the area covered by the
barrel RPC.  

A precise calibration of the RPC sub-detector is needed before it can
be used for time-of-flight measurements.
The aim of the calibration is to develop a procedure that transforms
data read out from a TDC channel into a flight-time of particles from
the production point in the target to the RPC pad feeding that
particular TDC channel. 
 An {\it in-situ} calibration procedure has been developed~\cite{rpc_ieee} 
 by using 
reconstructed charged particle tracks in the TPC from interactions 
produced by beams of positive particles with momenta of 3~\GeVc,
5~\GeVc, and 
8~\GeVc bombarding 0.05~$\lambda_{\mbox{\small{int}}}$ (nuclear
interaction length) thick Ta, Pb, Sn 
and Cu targets. Pion tracks were selected by using  $dE/dx$
measurement in the TPC. 
 The algorithm converts the TDC scale
into physical units, accounts for the arrival time of the beam
particle at 
the target, for the transit time of the signal within the pad
and from the pad to the preamplifier, and for temperature fluctuations
of the time response. It also corrects for so called {\it time-walk}
arising from the fact that signals with different pulse-height and
the same shape cross the fixed discriminator level at different time.
It is worth mentioning that the calibration procedure corrects for
 threshold crossing delays under the assumption of a
single, universal pulse-shape of the signals from all particles.
 The correction is quite large, up to 2 ns and is of the order of the
smallest time-of-flight to be measured.

More details of the design, layout, calibration and operational
parameters of the 
HARP RPCs can be found in~\cite{rpc_ieee}, \cite{rpc_first},
\cite{trans}, \cite{bogo}.

\section{RPC time response}
\label{sec:effect}

\begin{figure}
\centering
\includegraphics[width=3.in]{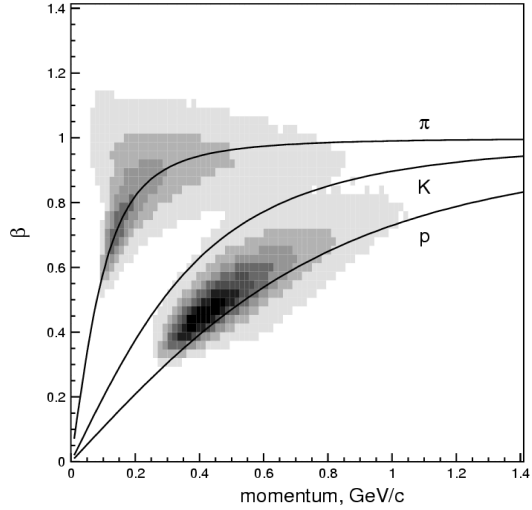}
\caption[] {\small {Relativistic velocity $\beta$, measured by the
    barrel ToF system, as a
    function of momentum, measured by the TPC, for positive
    particles. The
    curves represent theoretical dependence for pions (upper curve),
    kaons (middle curve) and protons (bottom curve). Proton signals are
    shifted towards higher $\beta$. }}
\label{pbeta}
\end{figure}

Figure~\ref{pbeta} demonstrates the
correlation between measured 
relativistic velocity $\beta$ of positively charged particles and
their momentum 
measured in the TPC when the {\it time-walk} is calibrated with pion
signals and applied to all tracks~\cite{rpc_ieee}.
It is clearly seen that proton signals are shifted with respect to the
theoretical curve. The rest of this section is devoted to the
understanding of the origin of this shift.

In Ref.~\cite{harpnimrebuttal} two mechanisms were suggested which
could be the cause of this effect. One possible explanation is the
fluctuation in  arrival time of the first cluster of the primary
ionization. This fluctuation is smaller for heavily ionizing
particles. The other possibility is a change of pulse shape near
threshold due to 
 a possibly different ionization and different gas
amplification regime.  

Since the observed rise-time of the signal to the discriminator
threshold is several ns,
a small difference in signal shape between heavily ionizing particles
(protons with momenta between  250~\MeVc and 800~\MeVc in our case)
and minimum ionizing particles (here pions with momenta above
150~\MeVc) could induce such an effect.
Effects below a ns are contained within the expected
avalanche formation time.

Dedicated analyses have been performed to understand the effect better.

\begin{figure}[tb]
\centering
\includegraphics[width=0.42\textwidth]{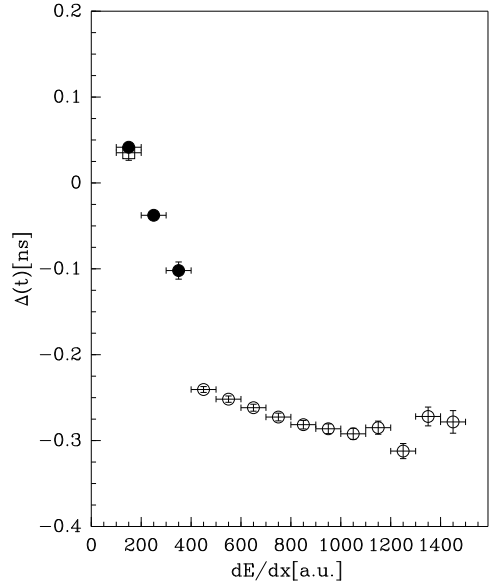}
\caption {Offset of the time-over-threshold of RPC signals selected 
  with measured total charge between 1400 and 1600 channel counts as a
  function of $dE/dx$ in the TPC gas.  The $dE/dx$ scale is in
  arbitrary units with the value of the minimum ionizing particle at
  $\approx 150$ counts.  The closed circles are obtained
  with pions, the open circles with protons using a $dE/dx$ selection,
  and the open square using protons with momentum above 1.5~\GeVc in
  addition to the  $dE/dx$ selection.
  The points where both pions and protons are minimum ionizing show a
  similar time offset.}
\label{fig:rpc:dedx}
\end{figure}

First, we compared the
relative time offset between the measured over-threshold time of the signal
and the predicted arrival time of protons above 1~\GeVc
with relativistic pions and with heavily ionizing protons as a function of
the measured energy loss of particles in the TPC gas 
(Figure~\ref{fig:rpc:dedx}).
To reduce the possible effect of the {\it time-walk}
  correction we use  a total
  pulse charge range  $1400<Q<1600$ in terms
  of channel counts, which is a region of relatively high pulse charge
  where still a large statistics can be obtained.
The energy loss in the TPC gas is  correlated with the
ionization density of the same particle  crossing the RPC 
gas gaps, just shifted by the material between the
TPC gas volume and the RPC gas gaps. Figure~\ref{fig:rpc:dedx} shows a
clear dependence of the time offset on 
the ionization density.
The points where both pions and protons are minimum ionizing show a
similar time offset, while the points at higher $dE/dx$ display a
trend justifying our hypothesis.
 
This result came as first evidence that a dependence on the 
ionization density in the gas gaps plays a role~\cite{harpnimrebuttal}.

\begin{figure}
\centering
\includegraphics[width=0.45\columnwidth]{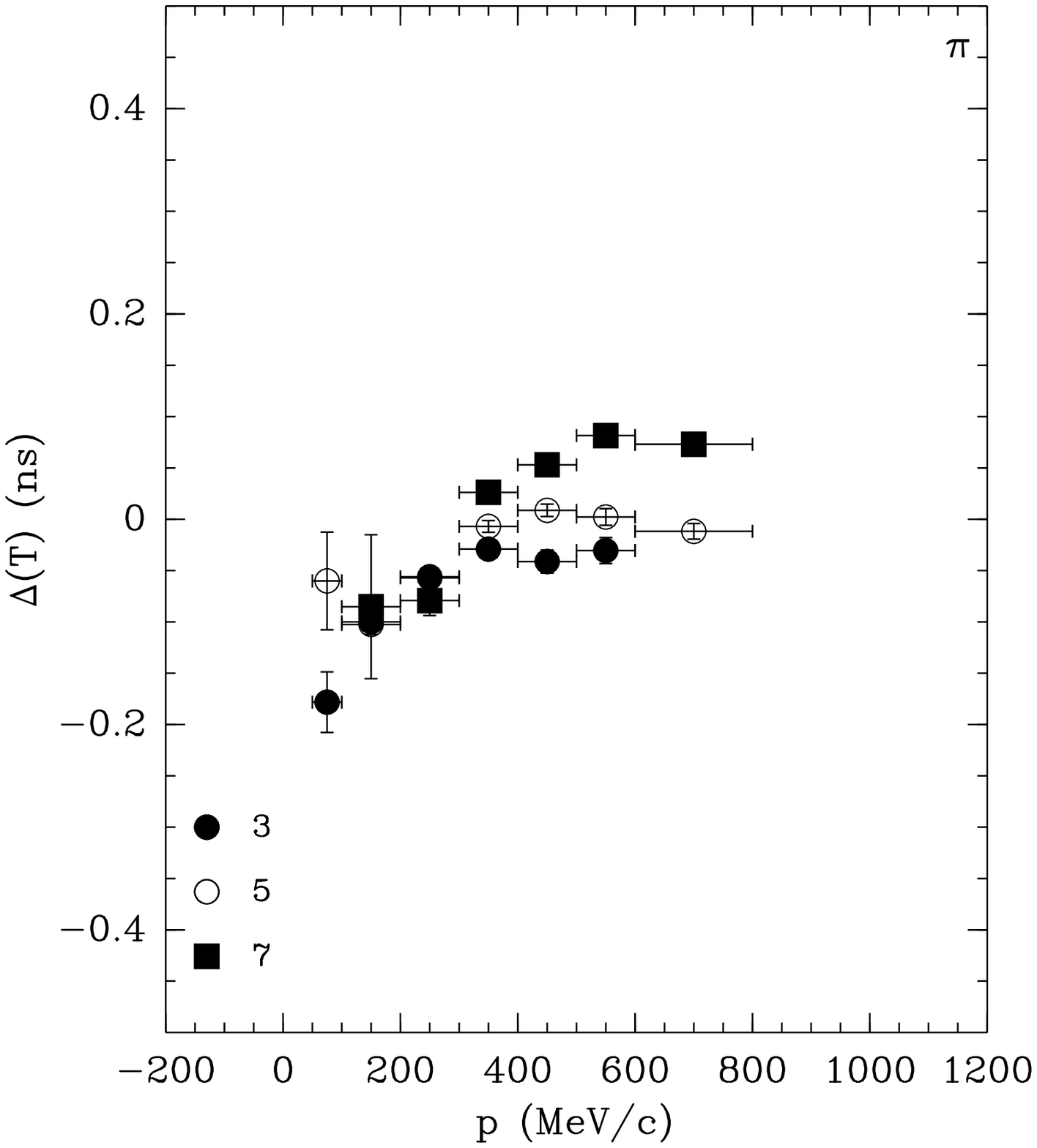}
\includegraphics[width=0.45\columnwidth]{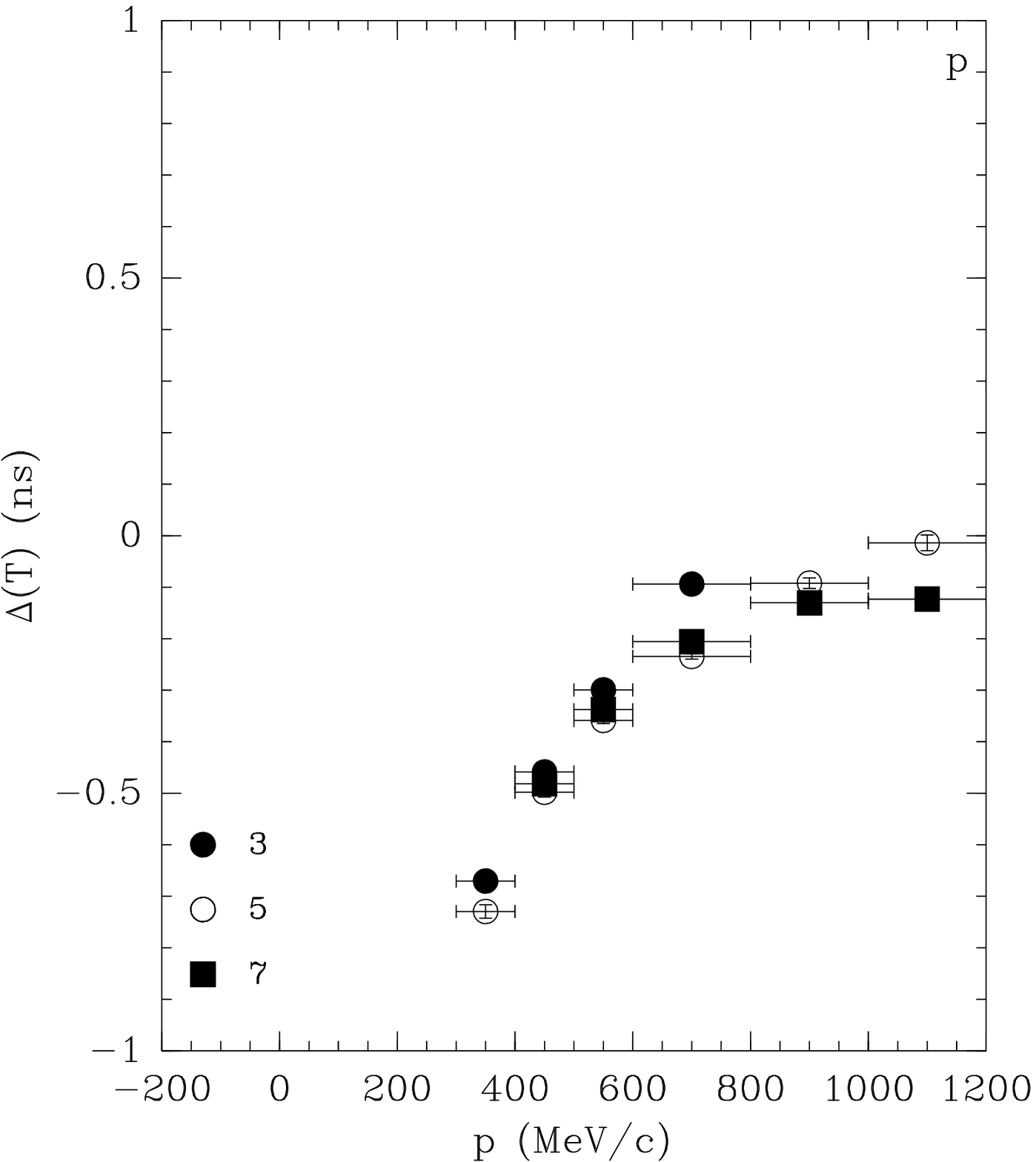}
\caption[] {\small {Difference of the measured time offset from the
  expected time offset for pions (left panel)
and protons (right panel) as function of the momentum measured in the
  TPC, for pad-rings 3, 5, and 7.}}
\label{dt_vs_p}
\end{figure}

Further, if our hypothesis is correct, the time difference between the
observed signal (corrected for the {\it time--walk}
per pad) and the expected arrival time calculated on the basis of the
track length and particle momentum as measured in the TPC 
should show a dependence as a
function of momentum for the same set of pad-rings due to the correlation 
between the energy loss and momentum for a given
particle species~\footnote{It is important to distinguish between
  different pad-rings 
  because particles emerging from the target cross the pad-rings at
  different inclinations, hence they produce different total
  ionization even having the same momentum.}.
Such a dependence was suggested in Ref.~\cite{rpc_ieee} 
and the respective plots are reproduced here in figure~\ref{dt_vs_p}.
The behavior seen in the figure indeed demonstrates a clear dependence
on ionization density ($dE/dx$) and a dependence on different path 
lengths at fixed $dE/dx$ only for low values of this quantity.

It should be noted that the above momentum dependencies were obtained
using the momentum of the particles measured in the TPC gas.
In Ref.~\cite{comments_nim}, \cite{comments_ieee} it has been pointed out
that a similar behavior can be obtained when a
systematic shift in the measurement of momentum is present.
Besides our confidence, based on various independent calibration
methods, that our momentum measurement is 
unbiased~(\cite{harpnim}, \cite{harpnimrebuttal},
\cite{harp:tantalum}, \cite{harpmomentum}) 
the effect of any momentum measurement bias in the TPC can be eliminated by an
analysis employing the kinematics of elastic scattering using a liquid
hydrogen target.
Such a measurement makes it possible to send a ``controlled beam'' of
slow protons through the TPC and towards the RPC system without the need to
measure the momentum of the recoil proton with the TPC.
The prediction for the momentum and direction of the recoil proton can
be obtained from the kinematics of the event by measuring the scattering
angle of the forward scattered proton or pion. 
Exploiting this, we have used an
exposure of the HARP detector where a 5~\GeVc beam of
protons and pions is directed onto a 60~\mm long liquid hydrogen
target. 
Elastic events are selected from the total sample of triggers requiring
one and only one track in the forward spectrometer, and exactly one
track in the TPC.  
The momentum of the track measured in the forward dipole spectrometer is
required to be consistent with elastic scattering hypothesis.
Only events with exactly one RPC hit are
retained.  
The position of the RPC hit has to be consistent with the predicted
impact point of the recoil track, using the direction of the forward
scattered track to define the trajectory of the recoil particle. 
This selection of elastic scattering events has a purity of
99\%~\cite{harp:tantalum}. 
Due to the acceptance of the forward spectrometer, the distribution of
selected recoil protons peaks at $74^\circ$ with respect to the beam
direction, thus almost perpendicular to the RPC chambers, with most of
the tracks contained between $60^\circ$ and $80^\circ$.
The recoil momenta are between 350~\MeVc and 600~\MeVc.
The geometry of the RPC system is such that the large majority of
selected recoil protons traverses pad-ring 3,
therefore, this measurement is performed using this pad-ring only.
In order to predict the time-of-flight of the particle reliably, the
initial momentum and energy loss in the material traversed has to be
described accurately.  
For example, a proton with 325~\MeVc momentum in the TPC gas 
has lost  in terms of momentum on average $\sim$~45~\MeVc with an
r.m.s.  of 15~\MeVc  
in the material 
between the interaction point and the TPC gas volume
and is going to lose $\sim$~83~\MeVc more (with an r.m.s. of 25~\MeVc)
in the outer 
field cage before reaching the RPC detection volume.
At 575~\MeVc the total reduction in momentum is $\sim$~38~\MeVc on
average with similar r.m.s.\footnote{The general HARP
  Monte Carlo simulation package (see~\cite{harpnim}) based on 
GEANT4 toolkit~\cite{geant4} has been used throughout the analysis
presented in this article.}

\begin{figure}
\centering
\includegraphics[width=0.49\columnwidth]{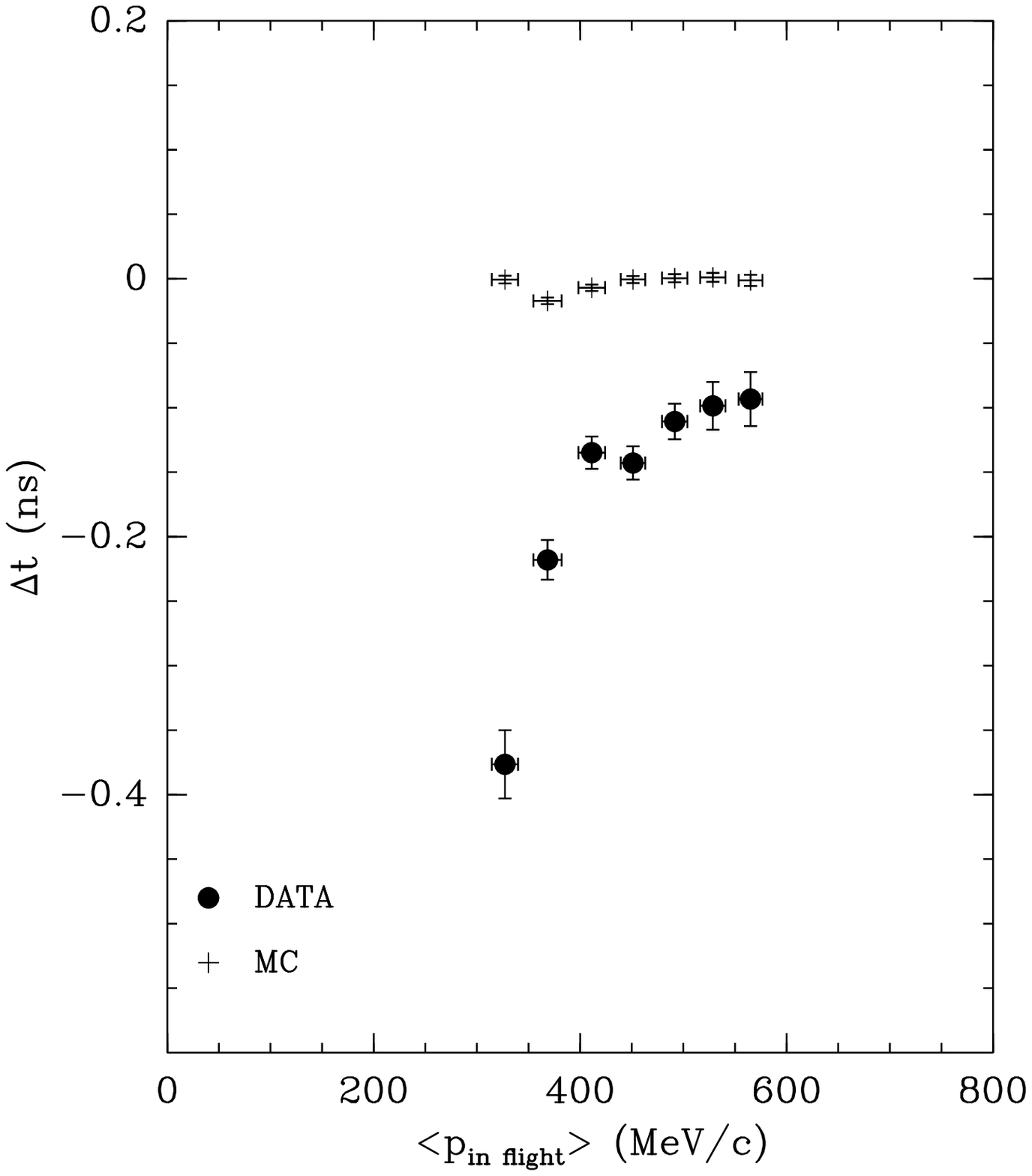}
\includegraphics[width=0.49\columnwidth]{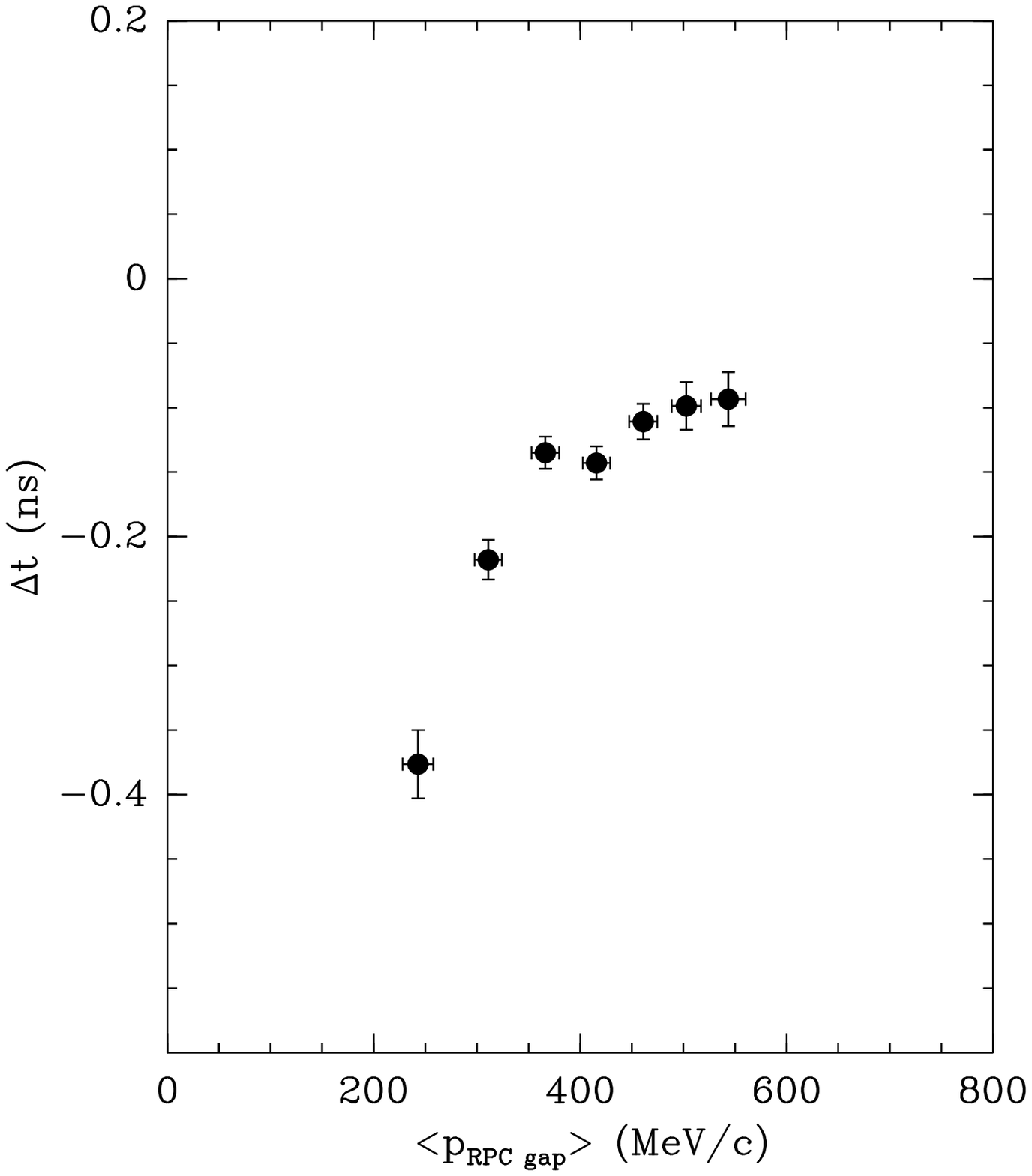}
\caption[] {\small {{\it Left panel:} The difference of the time offset measured
 in pad-ring 3 from the  expected time offset for protons as a
 function of the 
momentum along its 
 flight path  (in the gas volume of the TPC). The filled circles
 show the results of measurements using elastic scattering on hydrogen, the
 points without marker represent the simulation of the  measurement using the same
 reconstruction procedure. 
 The momentum was predicted using the kinematics of elastic scattering.
 Consistency of the simulated time difference with zero shows
 that the prediction of the flight time (and thus of the momenta) using
 the elastic scattering kinematics and Monte Carlo
 corrections in the reconstruction procedure for respective energy
 losses are correct.
 
 {\it Right panel:} The difference of the time offset measured
 in pad-ring 3 from the
 expected time offset for protons as a function of the momentum in the RPC
 gap.
}}
\label{fig:elas:dt:diff}
\end{figure}

The properties of the kinematics of elastic scattering are used to
predict the path-length of the particle from the interaction point to
the RPC detector and the time-of-flight over this distance.
Monte Carlo simulation
 is used to verify that the predicted and reconstructed
time and path-length agree within 5~\ps and 2~\mm, respectively. 
In the left panel of figure~\ref{fig:elas:dt:diff} the time difference,
defined as {\em the measured time minus the predicted time}, is
displayed as a 
function of the predicted momentum of the proton in the gas of the TPC. 
The momentum was predicted using the kinematics of elastic scattering.
The fact that the simulated difference of prediction and measurement is
consistent with zero shows that the prediction of the flight-time (and
thus of the momenta) using the elastic scattering kinematics and
Monte Carlo corrections in the reconstruction
procedure for respective energy losses  is
correct. The data exhibits a clear deviation pointing
to a difference in RPC time response as a function of the
momentum.
The right panel of figure~\ref{fig:elas:dt:diff} shows the same time
difference as a function of the momentum
predicted at the RPC detection volume.
Plots in figure~\ref{fig:elas:dt:diff} clearly demonstrate the existence
of a remarkable dependence of the RPCs response on the 
primary ionization density produced in the gas-gap by the particles of 
different types.  
The magnitude of the effect is in agreement with our previous 
analysis~\cite{harpnim}, \cite{rpc_ieee}, \cite{harpnimrebuttal}.  

\begin{figure}
\centering
\includegraphics[width=0.49\columnwidth]{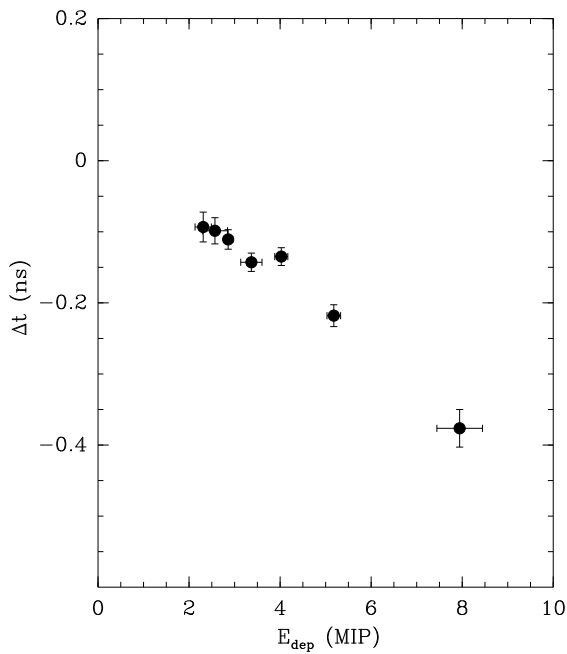}
\caption[] {\small {The difference of the time offset measured
 in pad-ring 3 from the
 expected time offset for elastically produced recoil protons as a
 function of the predicted 
 energy  deposition in the RPC gaps. 
 The energy deposition is expressed in units of the average energy
 deposition of a minimum ionizing particle (MIP).
}}
\label{fig:elas:dt:edep}
\end{figure}

In fact, one would expect that the RPC response does not depend directly
on momenta of detected particles, but only on their energy deposition in the gap.
Given the small range of path lengths in the RPC gas gaps  in {\em the
sample of elastic events}
it is natural to show the time differences as a function of total
energy deposition in the RPC gas.
The energy deposition can be predicted as a function of momentum using
our simulation. 
In figure~\ref{fig:elas:dt:edep} the same time difference measurements
as the ones of figure~\ref{fig:elas:dt:diff}
are displayed as a function of the above energy deposition. 
The range of this quantity in the data is between two and eight MIP (the
average deposition of one minimum ionizing particle).
In this limited range a nearly linear behavior of the shift in time
response is observed, with the extrapolation to one~MIP consistent with a
vanishing shift.
This is to be expected, since the RPC system was calibrated using pions
with momenta close to the value where their energy deposition is minimal.
Although the dependence does not show a flattening at higher energy
deposition in the measured range, one would expect this to occur at very
high ionization losses.

\section{Discussion}
\label{sec:discus}
We have clearly shown in the previous section that the time response
of RPCs with thin gas gaps depends on ionization density of detected
particles in those gaps even after correction for the usual {\it
  time-walk} based on the total charge of the signal. Evidence for
such an effect was reported by us earlier~\cite{harpnim}, \cite{rpc_ieee},
 \cite{harpnimrebuttal}, \cite{rebuttal_ieee}. 
Now we support our findings by strong
additional evidence
exploiting proton--proton and pion--proton elastic scattering kinematics
that allows us to avoid any dependence on the momentum measurement in the
TPC.

In view of the evidence presented in this work, the comments of Ammosov
et al. about our first indication~\cite{harpnim} of the effect,
e.g.~\cite{comments_nim}: 
\begin{quotation}
The timeslewing correction for the timing of pulses with
finite rise time has been well understood for over half a
century. The discovery that on top of the timeslewing
correction a further correction for protons 'due to their
higher ionization rate' is needed, is only one example out of
many grave misconceptions and mistakes contained in the
above-cited paper.
...
The above-mentioned discovery that the proton timeslewing
correction is different from the pion timeslewing
correction by 500 ps is a plain mistake stemming from the
lack of understanding of the interplay between TPC track
and RPC timing reconstruction.
\end{quotation} 
show that for the authors of~\cite{comments_nim} such an effect was
unexpected and therefore missed. 

It is
interesting to mention that, after more evidence for the discussed effect
was published in our papers~\cite{rpc_ieee}, \cite{harpnimrebuttal}, the same
authors acknowledged the effect without proper reference to our
work and tried to 
construct a physical model explaining it~\cite{dydak_rpc}. Their
estimation of 80~\ps time difference for protons at 400~\MeVc is
by factor of 2.5 smaller than our measurement.  
In a very recent preprint~\cite{dydak_tpc} the same authors exploit
the time-of-flight of low momentum protons measured by the same RPC system
as a benchmark for their procedure of momentum measurement in the HARP
TPC. They do not account for the effect reported here risking a
shift in their momentum calibration.

\section{Conclusions}
\label{sec:concl}

It was found that the time response of the thin gap RPC system used in the
HARP experiment and calibrated with the assumption of a unique
pulse-shape appeared to be 
different for pions and protons in the momentum range below 1~\GeVc.
In earlier publications we proposed the higher energy loss of protons as
a probable explanation of the effect.
In this paper it was clearly confirmed that the effect is present and
can indeed be
understood as originating from the considerably higher ionization
produced by the protons in the RPC.
Momentum measurement biases in the TPC, if any, have been eliminated
as possible cause of the effect.  

It should be noted that possible measurement of the effect using a
direct beam of protons with momenta as low as 300~\MeVc--600~\MeVc
would be interesting. However, it is
not an easy task due to the high energy losses of such protons.

\section*{Acknowledgments} 

We gratefully acknowledge the help and support of 
our colleagues from the HARP collaboration 
M.~Apollonio,
A. Bagulya,
G.~Barr,
F.~Bobisut,
M.~Bonesini,
S.~Bunyatov,
J.~Burguet--Castell,
C.~Buttar,
M.~Chizhov,
L.~Coney,
A.~De~Santo,
E.~Di~Capua,
U.~Dore,
J.~Dumarchez,
R.~Edgecock,
M.~Ellis,
R.~Engel,
F.~Ferri,
G.~Giannini,
D.~Gibin,
S.~Gilardoni,
P.~Gorbunov,
C.~G\"{o}\ss ling,
V.~Grichine,
P.~Gruber,
P.~Hodgson,
L.~Howlett,
I.~Kato,
A.~Kayis-Topaksu,
M.~Kirsanov,
J. Mart\'{i}n--Albo,
G.~B.~Mills,
M.C.~Morone,
P.~Novella,
D.~Orestano,
M.~Paganoni,
F.~Paleari,
V.~Palladino,
I.~Papadopoulos,
F.~Pastore,
S.~Piperov,
N.~Polukhina,
G.~Prior,
D.~Schmitz,
F.J.P.~Soler,
M.~Sorel,
A.~Tonazzo,
L.~Tortora,
G.~Vidal--Sitjes
and 
P.~Zucchelli.
We are indebted to
the PS beam staff
and to the numerous technical collaborators who contributed to the
detector design, construction, commissioning and operation.  
In particular, we would like to thank
G.~Barichello,
R.~Brocard,
K.~Burin,
V.~Carassiti,
F.~Chignoli,
D.~Conventi,
G.~Decreuse,
M.~Delattre,
C.~Detraz,  
A.~Domeniconi,
M.~Dwuznik,   
F.~Evangelisti,
B.~Friend,
A.~Iaciofano,
I.~Krasin, 
D.~Lacroix,
J.-C.~Legrand,
M.~Lobello, 
M.~Lollo,
J.~Loquet,
F.~Marinilli,
J.~Mulon,
L.~Musa,
R.~Nicholson,
A.~Pepato,
P.~Petev, 
X.~Pons,
I.~Rusinov,
M.~Scandurra,
E.~Usenko,
and
R.~van der Vlugt,
for their support in the construction of the detector.
The authors acknowledge the major contributions and advice of
M.~Baldo-Ceolin, 
M.T.~Muciaccia and A. Pullia
during the construction of the experiment.
The authors are indebted to
V.~Ableev,
P.~Arce,   %DR
F.~Bergsma,
P.~Binko,
E.~Boter,
C.~Buttar,  %DR
M.~Calvi, 
M.~Campanelli, %DR
C.~Cavion, 
A.~Chukanov, 
M.~Doucet,
D.~D\"{u}llmann,
R.~Engel,   %DR
V.~Ermilova, 
W.~Flegel,
A.~Grant,
P.~Gruber,   %DR
Y.~Hayato,
P.~Hodgson,  %DR
A.~Ichikawa,
A.~Ivantchenko,
I.~Kato,  %DR
O.~Klimov,
T.~Kobayashi,
A.~Krasnoperov,
D.~Kustov,
M.~Laveder,  
M.~Mass,
H.~Meinhard,
T.~Nakaya,
K.~Nishikawa,
M.~Paganoni,     %DR
F.~Paleari,  %DR
M.~Pasquali,
J.~Pasternak,   %DR
C.~Pattison,    %DR
M.~Placentino,
S.~Robbins,   %DR
S.~Sadilov, 
G.~Santin,  %DR
V.~Serdiouk,
S.~Simone,
V.~Tereshchenko,
A.~Tornero,   %DR
S.~Troquereau,
S.~Ueda, 
A.~Valassi,
F.~Vannucci,   %DR
R.~Veenhof
and
K.~Zuber   %DR
for their contributions to the experiment and to P. Dini for his
contribution to MC production.
We acknowledge the contributions of 
V.~Ammosov,
G.~Chelkov,
D.~Dedovich,
F.~Dydak,
M.~Gostkin,
A.~Guskov, 
D.~Khartchenko, 
V.~Koreshev,
Z.~Kroumchtein,
L.~Linssen, 
A.~De~Min,    
I.~Nefedov,
A.~Semak, 
J.~Wotschack,
V.~Zaets and
A.~Zhemchugov
to the work described in this paper.

The experiment was made possible by grants from
the Institut Interuniversitaire des Sciences Nucl\'eair\-es and the
Interuniversitair Instituut voor Kernwetenschappen (Belgium), 
Ministerio de Educacion y Ciencia, Grant FPA2003-06921-c02-02 and
Generalitat Valenciana, grant GV00-054-1,
CERN (Geneva, Switzerland), 
the German Bundesministerium f\"ur Bildung und Forschung (Germany), 
the Istituto Na\-zio\-na\-le di Fisica Nucleare (Italy), 
INR RAS (Moscow) and the Particle Physics and Astronomy Research Council (UK).
We gratefully acknowledge their support.
This work was supported in part by the Swiss National Science Foundation
and the Swiss Agency for Development and Cooperation in the framework of
the programme SCOPES - Scientific co-operation between Eastern Europe
and Switzerland.

% (used to reserve space for the reference number labels box)

\end{document}